\documentstyle[12pt,world_sci]{article}

\def\slsh#1{\slash \!\!\! #1}

\newcommand{\beqn}{\begin{eqnarray}}
\newcommand{\eeqn}{\end{eqnarray}}

\def\fb{{\rm fb}}

\def\abs#1{\left|#1\right|}
\def\tev{{\rm TeV }}
\def\gev{{\rm GeV }}
\def\Mev{{\rm MeV }}

\def\mev{{\rm meV }}

\def\spose#1{\hbox to 0pt{#1\hss}}
\def\lsim{\mathrel{\spose{\lower 3pt\hbox{$\mathchar"218$}}
     \raise 2.0pt\hbox{$\mathchar"13C$}}}
\def\gsim{\mathrel{\spose{\lower 3pt\hbox{$\mathchar"218$}}
     \raise 2.0pt\hbox{$\mathchar"13E$}}}
\def\simpropto{\mathrel{\spose{\lower 3pt\hbox{$\mathchar"218$}}
     \raise 2.0pt\hbox{$\propto$}}}

\def\beq{\begin{equation}}
\def\eeq{\end{equation}}
\def\barr{\begin{array}}
\def\earr{\end{array}}
\def\and{\qquad {\rm and } \qquad}
\def\etal{ {\it et al.} }

\def\PRD#1#2#3{{\sl Phys. Rev.} {\bf D#1}, #2 (#3)}

\def\PREP#1#2#3{{\sl Phys. Rep.} {\bf #1}, #2 (#3)}
\def\NPB#1#2#3{{\sl Nucl. Phys.} {\bf B#1}, #2 (#3)}

\def\tev{{\rm TeV }}
\def\gev{{\rm GeV }}
\def\mev{{\rm MeV }}

\def\abs#1{\left|#1\right|}

\def\tanb{\tan\beta}

\def\half{{1\over 2}}

\catcode`\@=11
\long\def\@makefntext#1{
\protect\noindent \hbox to 3.2pt {\hskip-.9pt  
$^{{\ninerm\@thefnmark}}$\hfil}#1\hfill}                

\def\@makefnmark{\hbox to 0pt{$^{\@thefnmark}$\hss}}  
        
\def\ps@myheadings{\let\@mkboth\@gobbletwo
\def\@oddhead{\hbox{}
\rightmark\hfil\ninerm\thepage}   
\def\@oddfoot{}\def\@evenhead{\ninerm\thepage\hfil
\leftmark\hbox{}}\def\@evenfoot{}
\def\sectionmark##1{}\def\subsectionmark##1{}}

\renewcommand{\thefootnote}{\fnsymbol{footnote}}

\newcounter{sectionc}\newcounter{subsectionc}\newcounter{subsubsectionc}
\renewcommand{\section}[1] {\vspace*{0.6cm}\addtocounter{sectionc}{1} 
\setcounter{subsectionc}{0}\setcounter{subsubsectionc}{0}\noindent 
        {\normalsize\bf\thesectionc. #1}\par\vspace*{0.4cm}}
\renewcommand{\subsection}[1] {\vspace*{0.6cm}\addtocounter{subsectionc}{1} 
        \setcounter{subsubsectionc}{0}\noindent 
        {\normalsize\it\thesectionc.\thesubsectionc. #1}\par\vspace*{0.4cm}}
\renewcommand{\subsubsection}[1]
{\vspace*{0.6cm}\addtocounter{subsubsectionc}{1}
        \noindent {\normalsize\rm\thesectionc.\thesubsectionc.\thesubsubsectionc. 
        #1}\par\vspace*{0.4cm}}

\newcounter{appendixc}
\newcounter{subappendixc}[appendixc]
\newcounter{subsubappendixc}[subappendixc]

\renewcommand{\appendix}[1] {\vspace*{0.6cm}
        \refstepcounter{appendixc}
        \setcounter{figure}{0}
        \setcounter{table}{0}
        \setcounter{equation}{0}
        \renewcommand{\thefigure}{\Alph{appendixc}.\arabic{figure}}
        \renewcommand{\thetable}{\Alph{appendixc}.\arabic{table}}
        \renewcommand{\theappendixc}{\Alph{appendixc}}
        \renewcommand{\theequation}{\Alph{appendixc}.\arabic{equation}}
        \noindent{\bf Appendix \theappendixc #1}\par\vspace*{0.4cm}}

\def\abstracts#1{{
        \centering{\begin{minipage}{12.2truecm}\footnotesize\baselineskip=12pt\noindent
        \centerline{\footnotesize ABSTRACT}\vspace*{0.3cm}
        \parindent=0pt #1
        \end{minipage}}\par}} 

\renewenvironment{thebibliography}[1]
        {\begin{list}{\arabic{enumi}.}
        {\usecounter{enumi}\setlength{\parsep}{0pt}
\setlength{\leftmargin 1.25cm}{\rightmargin 0pt}
         \setlength{\itemsep}{0pt} \settowidth
        {\labelwidth}{#1.}\sloppy}}{\end{list}}

\newcounter{itemlistc}
\newcounter{romanlistc}
\newcounter{alphlistc}
\newcounter{arabiclistc}

\newcommand{\fcaption}[1]{
        \refstepcounter{figure}
        \setbox\@tempboxa = \hbox{\footnotesize Fig.~\thefigure. #1}
        \ifdim \wd\@tempboxa > 6in
           {\begin{center}
        \parbox{6in}{\footnotesize\baselineskip=12pt Fig.~\thefigure. #1}
            \end{center}}
        \else
             {\begin{center}
             {\footnotesize Fig.~\thefigure. #1}
              \end{center}}
        \fi}

\newcommand{\tcaption}[1]{
        \refstepcounter{table}
        \setbox\@tempboxa = \hbox{\footnotesize Table~\thetable. #1}
        \ifdim \wd\@tempboxa > 6in
           {\begin{center}
        \parbox{6in}{\footnotesize\baselineskip=12pt Table~\thetable. #1}
            \end{center}}
        \else
             {\begin{center}
             {\footnotesize Table~\thetable. #1}
              \end{center}}
        \fi}

\def\@citex[#1]#2{\if@filesw\immediate\write\@auxout
        {\string\citation{#2}}\fi
\def\@citea{}\@cite{\@for\@citeb:=#2\do
        {\@citea\def\@citea{,}\@ifundefined
        {b@\@citeb}{{\bf ?}\@warning
        {Citation `\@citeb' on page \thepage \space undefined}}
        {\csname b@\@citeb\endcsname}}}{#1}}

\newif\if@cghi
\def\cite{\@cghitrue\@ifnextchar [{\@tempswatrue
        \@citex}{\@tempswafalse\@citex[]}}
\def\citelow{\@cghifalse\@ifnextchar [{\@tempswatrue
        \@citex}{\@tempswafalse\@citex[]}}
\def\@cite#1#2{{$\null^{#1)}$\if@tempswa\typeout
        {IJCGA warning: optional citation argument 
        ignored: `#2'} \fi}}

 1
 1
 1

\font\ninerm=cmr9

\begin{document}
\begin{flushright}
MPI-PhT/96-12\\
May 1996
\end{flushright}

\vskip4cm

\centerline{\normalsize\bf INDIRECT SIGNALS OF SUSY IN
}
\baselineskip=16pt
\centerline{\normalsize\bf 
GAUGE BOSON PAIR PRODUCTION
AT LEP AND NLC
\footnote{
to be published in the Preceedings of the XXXI Rencontres de Moriond,
Les Arcs, March 16th-23rd, 1996}
}


\centerline{}
\centerline{}
\centerline{\footnotesize Ralf Hempfling}
\baselineskip=13pt
\centerline{\footnotesize\it Max-Planck-Institut f\"ur Physik,
Werner-Heisenberg-Institut,}
\baselineskip=12pt
\centerline{\footnotesize\it F\"ohringer Ring 6, 80805 Munich, Germany}
\centerline{\footnotesize E-mail: hempf@mppmu.mpg.de}
\vspace*{0.3cm}

\vspace*{6.9cm}
\abstracts{
We compute the dominant
one-loop radiative corrections to the
cross-section for $e^+ e^- \rightarrow V_1 V_2$
$(V_{1/2} = \gamma, Z, W^\pm$) in the MSSM.
We find that the genuine vertex corrections
are very small. The oblique corrections are
potentially large enough to be tested
at LEP-II or NLC.
However, the sensitivity is below the one from other
high precision electro-weak
measurements but can serve as a self-consistency check.
}
\clearpage 
\normalsize\baselineskip=15pt
\setcounter{footnote}{0}
\renewcommand{\thefootnote}{\alph{footnote}}
\section{Introduction}
\begin{figure}[h]
\vspace*{13pt}
\vspace*{1.5truein}      
\includegraphics{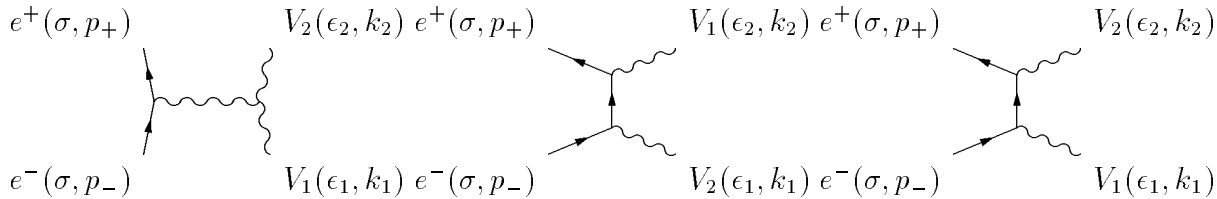}
\caption{
  Tree-level contributions to $e^+ e^- \rightarrow V_1 V_2$
}
\label{fig1}
\end{figure}

LEP-I experiments essentially
rule out new physics below $m_{\rm z}/2$~\cite{pdg}.
However, the high statistics allow even
to constrain virtual
effects from particles not kinematically accessible at LEP-I energy.
Thus, it is natural to ask whether this is also
possible at upcoming $e^+e^-$ colliders.
The higher center-of-mass energy, $E_{cm}$, opens new channels
even within the framework of the standard model.
In addition, the clean environment of an $e^+e^-$ collider is ideal
to test the triple gauge boson coupling (TGC).
In light of the success of the Standard Model (SM)
it is hard to imagine that the TGC will exhibit a
deviation from the SU(2)$\times$U(1) gauge structure.
Nonetheless, it might be possible to 
detect anomalies indicating virtuall effects
of new physics.
In table~\ref{tcollider} we have listed upcoming
collider experiments.
\begin{table}[h]
\tcaption{Upcoming $e^+e^-$ collider experiments}\label{tcollider}
$$
\begin{tabular}{lcrrl}
\hline
\hline
name & Type & $\sqrt{s}$ & $\int {\cal L} d t$ & date\\
\hline
LEP-II & circular & $165\sim 192~\gev$ & $4\times0.5~\fb^{-1}$ & now\\
NLC    & linear   & $0.5 \sim 2~\tev$ & $50\sim 200~\fb^{-1}$ &2005/10\\
\hline
\end{tabular}
$$
\end{table}
In $e^+e^-$ colliders a pair of gauge bosons can be produced
in three possible chanels as
depicted in fig.~\ref{fig1}.
Note, there is no s-chanel (no u-chanel)
for $\gamma\gamma$, $\gamma Z$, $ZZ$ ($W^+W^-$) production at tree-level.
In table~\ref{tv1v2} we have listed the cross-section for
gauge boson pair production in an $e^+ e^-$ collider 
obtained by integrating the differential cross-section
\beqn
{d \sigma\over d \cos \theta}(e^+e^-\rightarrow V_1 V_2)
= \frac{f_S \alpha^2_{\rm em}}{4s}\sqrt{1-\frac{4m_W^2}{s}}\abs{\cal A}^2\,,
\eeqn
where the symmetry factor $f_S = 1$ (1/2) for $W^+W^-$ and $\gamma Z$
($\gamma\gamma$ and $ZZ$). 
We also present the expected precision to which this cross-section can be measured
at LEP-II ($E_{cm}=200~\gev$)
and NLC ($E_{cm}=500~\gev$) assuming that the statistical errors dominate.
\begin{table}[h] 
\tcaption{cross-sections and expected statistical errors}\label{tv1v2}
$$
 \begin{tabular}{|l||r|r|r|r|r|r||r|r|} \hline
    \hline
    $\sqrt{s}$ [in GeV]& $165$ & $176$ &$190$
   &$205$ & $500$ & $2000$ & LEP-II& NLC\\ 
    \hline\hline
    $WW      $ & 10.7 & 16.1 & 17.7 & 17.5 & 6.5  & 0.82  &0.5
$\%$&0.2$\%$ \\
    $WW_{F-B}$ &  3.0 &  7.8 & 10.7 & 11.9 & 6.0  & 0.79  &0.7
$\%$&0.2$\%$ \\
    $ZZ$       &  -   &  -   &  1.0 &  1.2 & 0.37 & 0.045 &2.0
$\%$&0.7$\%$ \\
$\gamma Z$     & 21.1 & 16.6 & 12.7 &  9.9 & 1.1  & 0.064 &0.6
$\%$&0.4$\%$ \\
$\gamma\gamma$ & 5.8  &  5.1 &  4.4 &  3.8 & 0.63 & 0.039 &1.0
$\%$&0.6$\%$ \\
    \hline
  \end{tabular}
$$
\end{table}
With a precision in the percent range these processes are potentially
sensitive to radiative corrections (RC).
The agreement (disagreement) of the theoretical and 
experimental values provide an important self-consistency check
of the SM (hint for new physics).
However, before we can appreciate the importance of this process
we have to answer the question whether we
can actually learn anything new from testing the TGC?
One of the most important achievements at LEP-II
will be a very precise measurement of $m_{\rm w}$
to $\delta m_{\rm w}=\pm 50~\mev$.
This will pose the strongest constraint on new physics
via $\Delta r$\cite{M&S} with an expected precision
\beqn
\delta \Delta r &=& \pm 0.1\% (Th)
\pm 1\%{\delta m_{\rm w}\over 170~\Mev}
\pm 1\%{\delta m_{\rm t}\over 30~\gev}\,.
\nonumber
\eeqn
Here the uncertainty of the top
quark mass of presently $\delta m_{t}=9~\gev$ is
expected to reach $5~\gev$.
Similar constraints can be derived from measurements of the
forward-backward asymmetry, neutral current processes, and the running
$\alpha_{\rm em}$.
Thus, any virtuall effects of new physics on
$\sigma(e^+e^-\rightarrow V_1 V_2)$
has to be compared to the ones on $\Delta r$, etc.
In this paper we will focus on the
minimal supersymmetric extension of the SM (MSSM)\cite{susyreview}.

\section{RC due to Squarks/Sleptons}

The calculation of the RC to the gauge boson production
in $e^+e^-$ collider in the SM is quite elaborate\cite{RCeeww}\cite{RCeezz}.
Thus, in extending this calculation to the MSSM
we restrict ourselves to the dominant effects
expected to arise from the squark/slepton
sector which are enhanced by
\begin{itemize}
\item colour factor, $N_c = 3$ 
\item number of generations, $N_g = 3$ 
\item large top Yukawa coupling.
\end{itemize}
None of these enhancements exist for the
selectron-chargino loops.
Thus, we will rely throughout this work on the
assumption that these contribution can be neglected.
(For a more complete treatment see ref.~6).

\begin{figure}
\vspace*{13pt}
\vspace*{2.7truein}      
\includegraphics{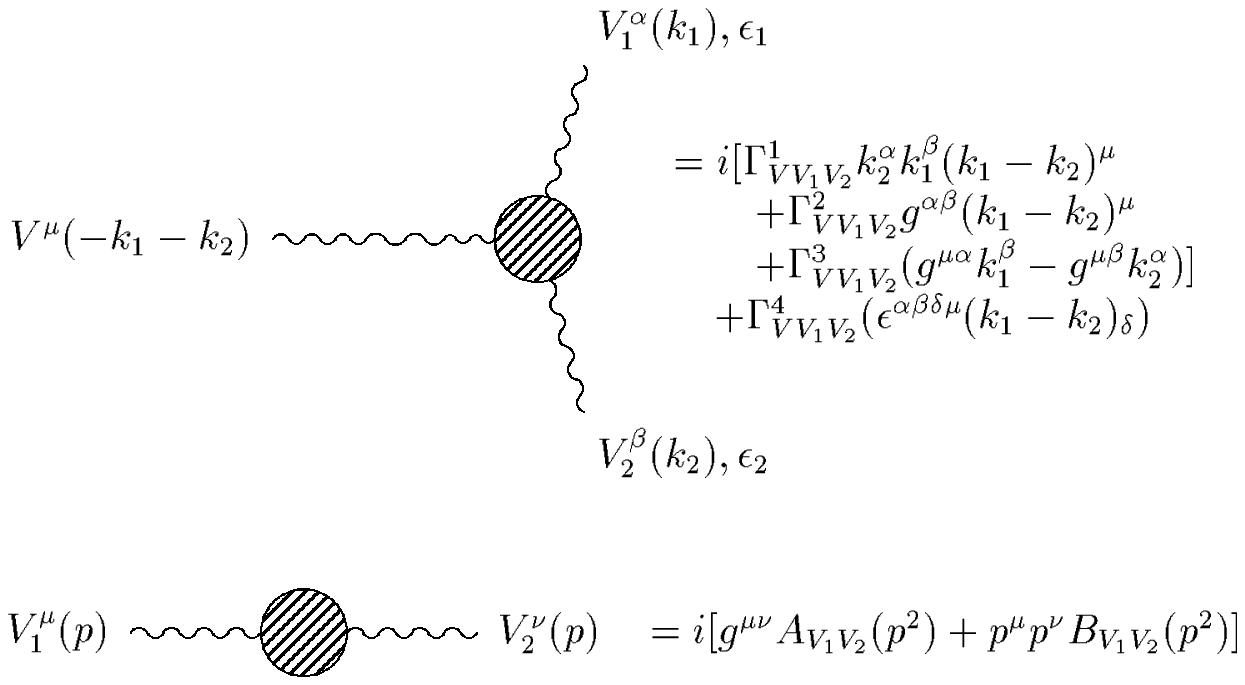}
\caption{
The Lorentz invariant decompositions of the gauge boson
self-energies and the TGC.
The ellips in the first diagram
stands for all contributions proportional to $k_1^\alpha$, $k_2^\beta$, $(k_1+k_2)^\mu, ...$.
}
\label{fig2}
\end{figure}
One consequence of our assumption is that virtuall SUSY contributions only
enter via the gauge boson self energy and the TGC. 
(In this sence, our approach is similar to ref.~7 where the supersymmetric RC to the TGC
were calculated). As a result, 
the amplitude can be written as ${\cal A}(\sigma,\epsilon_1,\epsilon_2,s,t)
 = \sum {\cal M}_i^\sigma(\epsilon_1,\epsilon_2) F_i^\sigma(s,t)$
with the matrix elements ($\sigma = \pm 1/2$; $P = 1/2+\sigma \gamma_5$)\cite{RCeeww}
\begin{eqnarray*}
    {\cal M}^\sigma_1(\epsilon_1,\epsilon_2) & 
= & \bar{v}(p_+)\slsh{\epsilon}_+^*(\slsh{k}_+-\slsh{p}_+)
    \slsh{\epsilon}_-^*P u(p_-) \\
    {\cal M}^\sigma_2(\epsilon_1,\epsilon_2) & 
= & \half \bar{v}(p_+)(\slsh{k}_+-\slsh{k}_-)
    (\epsilon ^*_+ \cdot \epsilon ^*_-)P u(p_-)\,,\\
    {\cal M}^\sigma_3(\epsilon_1,\epsilon_2) & 
= &\bar{v}(p_+)[\slsh{\epsilon} ^*_+ (\epsilon ^*_-
      \cdot k_+)-\slsh{\epsilon}^*_-(\epsilon^*_+ \cdot k_-)]P u(p_-)\,,\\
    {\cal M}^\sigma_4(\epsilon_1,\epsilon_2) & 
= &\bar{v}(p_+)[\slsh{\epsilon} ^*_+ (\epsilon ^*_-
      \cdot p_-)-\slsh{\epsilon}^*_-(\epsilon^*_+ \cdot p_+)]P u(p_-)\,,\\
    {\cal M}^\sigma_5(\epsilon_1,\epsilon_2) & 
= &\half\bar{v}(p_+)(\slsh{k}_+-\slsh{k}_-)
    P u(p_-)(\epsilon _+^* \cdot k_-)(\epsilon _-^* \cdot k_+)\,,\\
    {\cal M}^\sigma_6(\epsilon_1,\epsilon_2) & 
= &\bar{v}(p_+)[\slsh{\epsilon} ^*_+ (\epsilon ^*_-
      \cdot k_+)+\slsh{\epsilon}^*_-(\epsilon^*_+ \cdot k_-)]P u(p_-)
\label{melements}\,.\\
\end{eqnarray*}
The formfactors for $e^+e^- \rightarrow W^+W^-$ are
\beqn
F_1^{-1/2} \! &=& {g^2\over 2t}\left(1+2{\delta g\over g}
\right) -4{g_{eeZ}^L \Gamma_{ZW^+W^-}^4\over s-m_{\rm z}^2}
+4{e \Gamma_{\gamma W^+W^-}^4 \over s} \,,\nonumber\\
F_1^{+1/2} \! &=& 4{g_{eeZ}^R \Gamma_{ZW^+W^-}^4\over s-m_{\rm z}^2}
-4{e \Gamma_{\gamma W^+W^-}^4 \over s}\,,\nonumber\\
F_2^\sigma &=& 2{g_{eeZ}^\sigma g_{ZW^+W^-}\over s-m_{\rm z}^2}
\left(1+{\delta g_{eeZ}^\sigma\over g_{eeZ}^\sigma}
+{\delta g_{ZW^+W^-}\over g_{ZW^+W^-}}
+{A_{ZZ}(s)-A_{ZZ}(m_{\rm z}^2)\over s-m_{\rm z}^2}
-\delta Z_{ZZ} \right. \nonumber\\
&&\left. - \frac{e}{g_{ZW^+W^-}} {A_{\gamma Z}^{ren}(s)\over s}+
{\Gamma_{ZW^+W^-}^2+4\sigma \Gamma_{ZW^+W^-}^4 \over g_{ZW^+W^-}}
\right) \nonumber \\
&& - 2 {e^2\over s}
\left(1+2 {\delta e\over e}
+{A_{\gamma\gamma}(s)\over s}
-\delta Z_{\gamma\gamma} 
+\frac{g_{ZW^+W^-}}{e}\frac{A_{\gamma Z}^{ren}(s)}{s-m_Z^2}
+{\Gamma_{\gamma W^+W^-}^2
+4\sigma \Gamma_{\gamma W^+W^-}^4\over e}\right)\,,\nonumber\\
F_3^\sigma &=&-F_2^\sigma
-{\Gamma_{Z W^+W^-}^2+\Gamma_{Z W^+W^-}^3 +2\sigma \Gamma_{Z W^+W^-}^4\over g_{ZW^+W^-}}
-{\Gamma_{\gamma W^+W^-}^2+\Gamma_{\gamma W^+W^-}^3 +2\sigma \Gamma_{\gamma W^+W^-}^4\over e}
\,,\nonumber\\
F_4^\sigma &=& -4 \sigma {g_{eeZ}^\sigma \Gamma_{ZW^+W^-}^4\over s-m_{\rm z}^2}
-4 \sigma {e \Gamma_{\gamma W^+W^-}^4\over s}\,,\nonumber\\
F_5^\sigma &=&
-2{g_{eez}^{P} \Gamma_{WWZ}^1\over s-m_{\rm z}^2}+2\frac{e\Gamma_{WW\gamma}^1}{s}
~\qquad P =  L,R\,,
\eeqn
where
$g^{+1/2}_{eeZ} = g\sin^2 \theta_{\rm w} / \cos \theta_{\rm w}$,
$g^{-1/2}_{eeZ} = g^{+1/2}_{eeZ} + g/ 2\cos \theta_{\rm w}$,
$g^\sigma_{ZWW} =  g\cos \theta_{\rm w}$.
The formfactors for $e^+e^-\rightarrow V_1 V_2$ ($V_{1/2} = \gamma, Z$) are .
Note that for squark-loops the factors $\Gamma^4_{V W^+W^-}$ and
all $\Gamma^i_{V V_1 V_2}$ defined in fig.~2 vanish.

\section{Numerical Result}
\begin{figure}
\vspace*{13pt}
\vspace*{2.8truein}      
\includegraphics{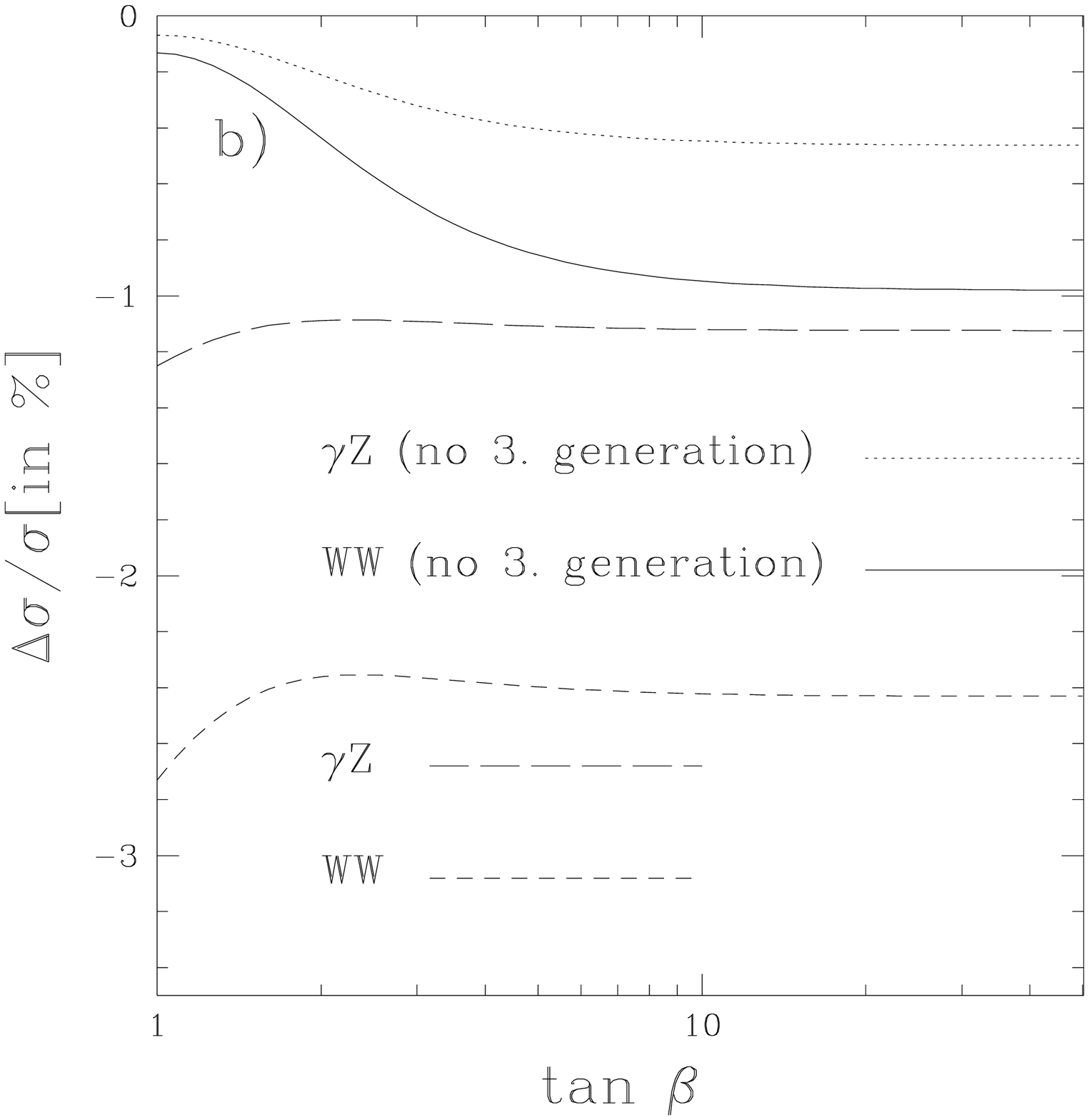}
\includegraphics{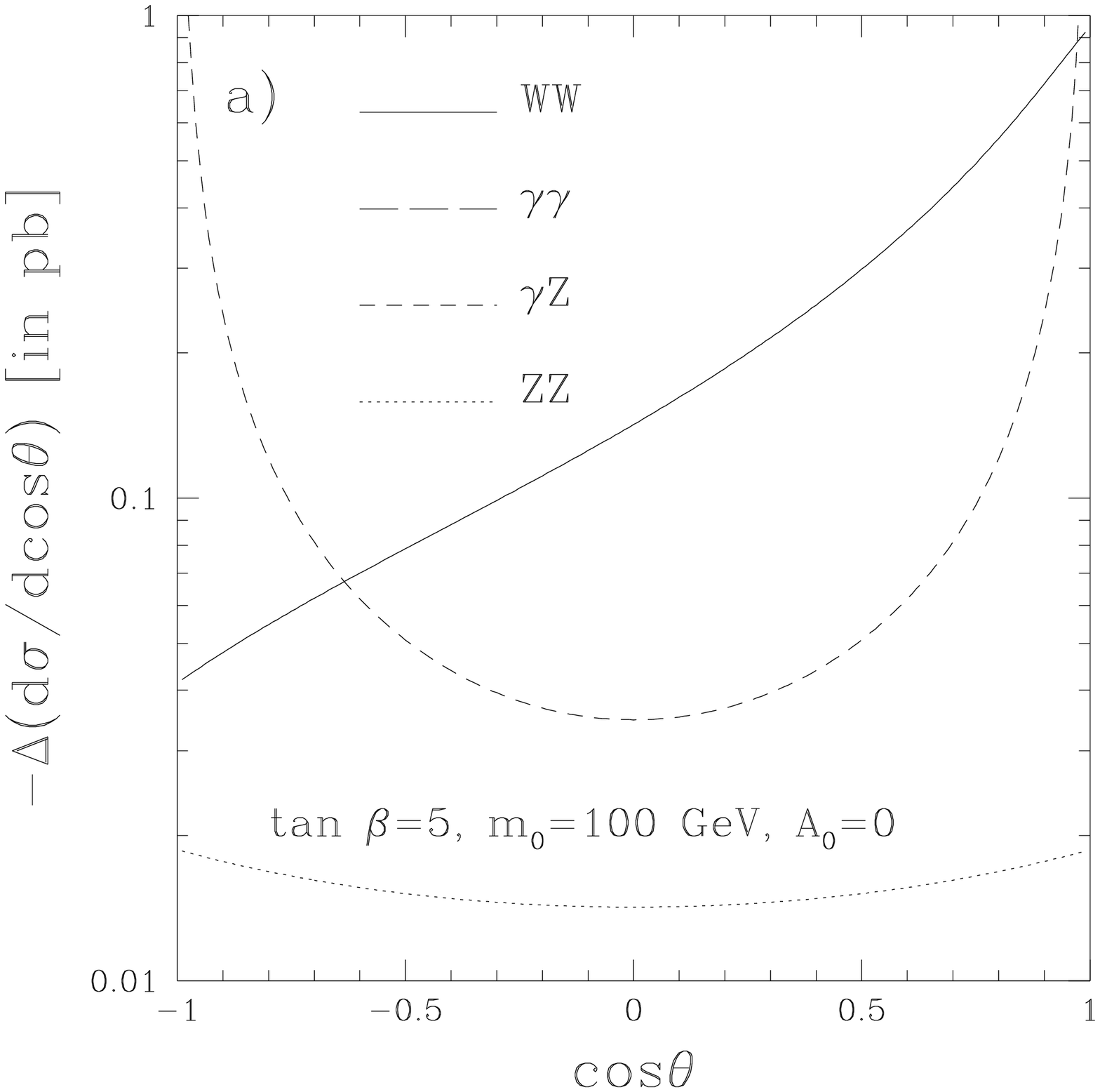}
\caption{differential and total cross-sections vs. $\cos \theta$ and $\tanb$,
respectively. In b) results are presented with and without the 3. generation
squarks for $E_{cm}=190~\gev$.
}
\label{fig3}
\end{figure}

Our sfermion mass spectrum is characterized by a universal
mass parameter $m_0$ (here, we omit the possibility of mass-splitting 
due to renormalization group evolution), a $L/R$ mixing parameter
$A_0$ and the ratio of Higgs VEVs $\tan \beta$.
In fig.~\ref{fig3} we see that the differential cross-sections with $\gamma$'s in the final
state diverges for $\abs{\cos \theta} = 1$. In this case, we choose the range of
integration as $\abs{\cos \theta} < 0.95$.
Furthermore, we find that the dominant RC come indeed from the 
third generation squarks
due to large mass splitting within SU(2) multiplet.
For the superpartners of the light fermions this splitting 
is generated by the SU(2)$_L$ $D$ term,
$M^2_{u_L} - M^2_{d_L} = g\langle D^3 \rangle = m_{\rm w}^2 \cos 2\beta$.
\begin{figure}
\vspace*{13pt}
\vspace*{2.8truein}      
\includegraphics{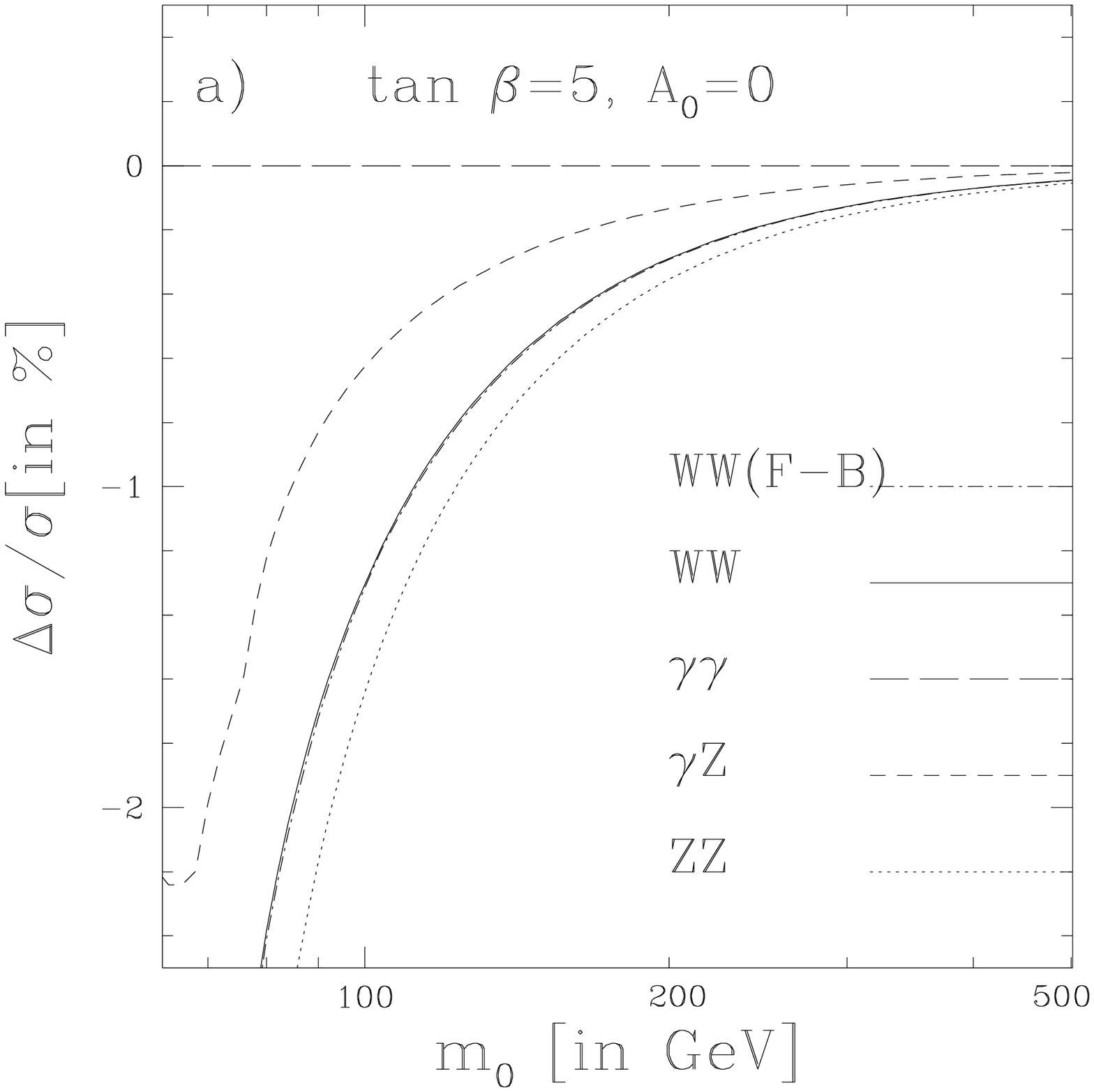}
\includegraphics{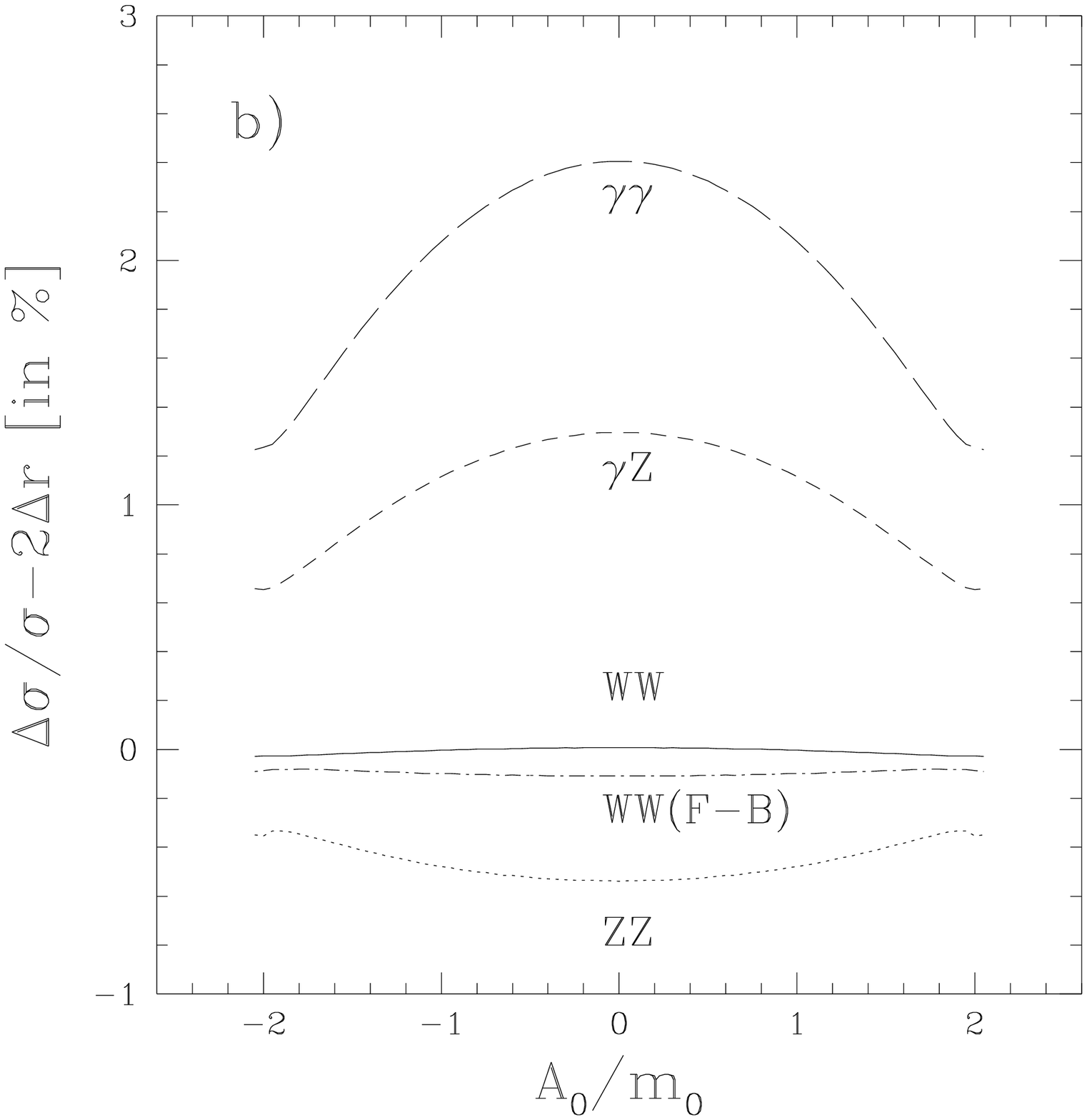}
\caption{
the sfermionic RC to the total cross-sections vs. $m_0$ and $A_0$.
The born term is parameterized by a) $\alpha_{\rm em}$ and b) $G_\mu$.
}
\label{fig4}
\end{figure}

The counter terms in eq.~2 depend on the renormalization scheme,
ie. the parameterization of the Born term. 
Sofar we have used $g^2 = {4 \pi \alpha_{\rm em}/\sin^2 \theta_{\rm w}}$
It is easy to understand that the RC to  $e^+e^-\rightarrow \gamma \gamma$
vanish in this scheme while those to
$e^+e^-\rightarrow W^+W^-$ can become quite large (see fig. 4b).
However, if we change the renormalized coupling constant to
$g^2 \rightarrow \sqrt{2} m_{\rm w}^2 G_\mu$ then we have to replace
$\delta g/g \rightarrow \delta g/g +  \Delta r$ and
$\Delta \sigma/\sigma \rightarrow \Delta \sigma/\sigma - 2 \Delta r$.
In this scheme the
RC to  $e^+e^-\rightarrow W^+W^-$
are very tiny (fig.~\ref{fig4}b).
For $\sigma(e^+e^-\rightarrow ZZ)$ 
and $\sigma(e^+e^-\rightarrow \gamma Z)$ the situation is similar.
In the SM it is convenient to use $G_\mu, \alpha_{\rm em}, m_{\rm z}$
and possibly $m_{\rm w}$ to parameterize the tree-level
term because these observables are know to such a high precision. 
However, in the MSSM the largest uncertainty arise
from our ignorance of the SUSY parameters.
Thus, it is convenient to parameterize
in such a way that RC cancel, even if that means to use observables with larger errors.
Eg.: it is easy to see that no sfermionic RC exist
to the relation
\beqn
{d \sigma\over d \cos \theta}(e^+e^-\rightarrow \gamma Z)
= \alpha_{\rm em} \Gamma(Z\rightarrow e^+e^-)\times ...\, ,\nonumber
\nonumber\eeqn
where the ellips stands for some kinematical factors.
Thus, with an error of $0.3\%$\cite{pdg}
in the leptonic width of the $Z$ boson
an expected precision of at best $0.4\%$ for
$\sigma(e^+e^-\rightarrow \gamma Z)$ at NLC
this process is not suited to yield new information
on sfermions.
A similar relation holds between 
$\sigma(e^+e^-\rightarrow Z Z)$,
$ \Gamma(Z\rightarrow e^+e^-)$, and
$ \Gamma(Z\rightarrow hadrons)$.
A deviation of the TGC from the SM prediction
(assuming $\Delta r$ agrees with the SM prediction)
will indicate new physics other than the MSSM.

\section{Summary}
We have investigated virtual effects of sfermions
on the cross-sections of gauge boson pair productions.
We find that
\begin{itemize}
\item
the genuine vertex corrections of sfermions to $ZW^+W^-$
and $\gamma W^+W^-$ are very small;
\item
the genuine vertex corrections of sfermions to $V_1V_2V_3$
vanish if $V_i = \gamma, Z$;
\item
all oblique corrections can be absorbed by
a suitable parametrization of the Born term;
\item
the experimental precision on $\sigma(e^+e^-\rightarrow V_1V_2)$
is lower than the precision of the corresponding 
EW observables $\Gamma(Z\rightarrow hadrons)$ or
$\Gamma(Z\rightarrow leptons)$ etc;
\item
sfermions cannot explain any anomalous TGC 
that may be observed at LEP-II unless there is also evidence of new physics
in other observables (e.g. $\Delta R$);
\item
if a deviation in e.g. $\Delta r$ is found
then $\sigma(e^+e^-\rightarrow V_1V_2$)
provides an additional cross-check;

\end{itemize}

\section{Acknowledgements}
I would like to express my gratitude
to my collaborators
B.A. Kniehl and E.A. Siemes for their
active participation in this project.
Furthermore, I would like to thank
the organizers of this workshop for
creating such a stimulating and inspiring atmosphere.

\end{document}